\documentclass[a4paper,10pt,pre,twocolumn,showpacs,aps,floats,superscriptaddress]{revtex4}

\usepackage{epsfig}

\begin{document}

\title{Local versus Global Knowledge in the Barab\'asi-Albert scale-free
network model} 

\author{Jes\'us G\'omez-Garde\~nes}

\affiliation{Departamento de Teor\'{\i}a y Simulaci\'on de Sistemas
  Complejos. Instituto de Ciencia de Materiales de Arag\'on.
  C.S.I.C. - Universidad de Zaragoza, Zaragoza 50009, Spain}

\affiliation{Instituto de Biocomputaci\'on y F\'{\i}sica de Sistemas
Complejos, Universidad de Zaragoza, Zaragoza 50009, Spain}

\author{Yamir Moreno}

\affiliation{Instituto de Biocomputaci\'on y F\'{\i}sica de Sistemas
Complejos, Universidad de Zaragoza, Zaragoza 50009, Spain}

\date{\today}

\widetext

\begin{abstract} 

The scale-free model of Barab\'asi and Albert gave rise to a burst of
activity in the field of complex networks. In this paper, we revisit
one of the main assumptions of the model, the preferential attachment
rule. We study a model in which the PA rule is applied to a
neighborhood of newly created nodes and thus no global knowledge of
the network is assumed. We numerically show that global properties of
the BA model such as the connectivity distribution and the average
shortest path length are quite robust when there is some degree of
local knowledge. In contrast, other properties such as the clustering
coefficient and degree-degree correlations differ and approach the
values measured for real-world networks.

\end{abstract}

\pacs{89.75.-k, 89.75.Fb, 05.70.Jk, 05.40.a}

\maketitle


During the last several years, many scientists have scrutinized the
world around us to unravel the complex patterns of interconnections
that characterize seemingly diverse social \cite{pnas}, biological
\cite{ref5,ref6} and technological systems \cite{ref7,ref8}. These
systems have been shown to exhibit common features that can be
captured using the tools of graph theory or in more recent terms,
network modeling. At the same time, network models of diverse kinds
have been proposed with the aim of describing and explaining the
properties of real webs \cite{doro,bara02}. It turns out that most
real networks are better described by growing models in which the
number of nodes (or elements) forming the net increases with time and
that the probability that a given node has $k$ connections to other
nodes follows a power-law $P_k\sim k^{-\gamma}$, with $\gamma \le
3$. Additionally, the study of processes taking place on top of these
networks has led us to reconsider classical results obtained for
regular lattices or random graphs due to the radical changes of the
system's dynamics when the heterogeneity of complex networks can not
be neglected \cite{book1,book2,av03,moreno03}.

The first scale-free network model, introduced by Barab\'asi and
Albert (BA), postulated that there are two fundamental ingredients of
many real networks \cite{bar99,bar99a}: their growing character and
the preferential attachment (PA) rule. The preferential attachment
rule considers that the probability that an old node links to newly
added nodes is proportional to its degree $k$. It summarizes the
common belief that the more rich you are, the more likely it is that
your richness grows, that's why the term rich-gets-richer has been
used to refer to the PA rule \cite{bar99a}. However, the BA model
assumes that one knows the connectivity of all nodes when a new node
links to the network. This is clearly an unrealistic assumption. This
drawback of the model construction has not passed unnoticed and many
models have been introduced to produce scale-free networks and to test
whether or not the basic assumptions of the BA recipe are necessary
conditions to build up these networks \cite{doro,bara02}. 

Growing models which produce scale-free graphs with arbitrary
$\gamma$-exponents, and non-random correlations can be found nowadays
in the scientific literature. On the other hand, there are some models
in which the PA rule is limited to a neighborhood due to geographic
constraints \cite{amaral}, or where its linear character is
investigated \cite{redner}. Recently, Caldarelli {\em et al.}
\cite{cald} have shown that one can produce SF networks without
assuming preferential attachment at all. As a byproduct, other
properties of the network fit well with those of real-world
graphs. They introduced an intrinsic fitness model in which two nodes
are connected with a probability that depends on their fitness. Note,
additionally, that the way in which the fitness parameter was
introduced is different from the model in \cite{bia1}.

In this paper, we adopt a different perspective. Our aim is to test to
what extend the global character of the PA rule in the original BA
model is important. We introduce a model in which the PA is applied
only to a neighborhood of the newly added node depending on the value
of a variable which measures the affinity between different nodes. By
going down from the BA limit of the model to the the limit where all
nodes are distinct, we test to what extend the global knowledge of
each node's connectivity is fundamental to get a scale-free
graph. Through numerical simulations we find that in a wide range of
the model parameters, average quantities such as the connectivity
distribution and the shortest path length are not affected by the use
of local knowledge of the network whereas other properties like the
clustering coefficient are more sensitive to local details.


Our model is defined in two layers. The first discriminates among all
the nodes by assigning to each node at the moment of its creation a
parameter $a_i$ which measures how close or distinct a given node is
from the rest of the elements that compose the network. Then, we apply
the preferential attachment rule in the neighborhood defined by nodes
with common affinities. Specifically, the network is constructed by
repeated iteration of the following rules: {\it i)} Start from a small
core of nodes, $m_o$, linked together. Assign to each of these $m_o$
nodes a random affinity $a_i$ taken form a probability
distribution. In what follows, we will use for simplicity a uniform
distribution between $(0,1)$; {\it ii)} At each time
step, a new node $j$ with a random affinity $a_j$ is introduced and
linked to $m$ nodes already present in the network according to the
rules specified below; {\it iii)} Search through all nodes of the
network verifying whether or not the condition $a_i-\mu \le a_j \le
a_i+\mu$ is fulfilled, where $\mu$ is a parameter that controls the
affinity tolerance of the nodes. The nodes that satisfy the affinity
condition are grouped in a set $A$ as potential candidates to gain new
links; {\it iv)} Apply the preferential attachment rule to the set $A$
\cite{note2}, {\em i.e.}, when choosing the nodes to which the new
vertex links, we impose that the probability that vertex $i$ connects
to the new node depends on its connectivity such that
\begin{equation}
\Pi(k_i)=\frac{k_i}{\sum_{s\in{A}}k_s}
\label{eq1}
\end{equation}
; and finally {\it v)} Repeat steps {\it ii-iv} such that the final
size of the network is $N=m_o+t$.

Thus, after $t$ time steps a network made up of $N$ nodes builds
up. It is worth mentioning that the inclusion of the affinity
parameter $a$ is not a mere artifact. Indeed, most real systems are
formed by non-identical elements and thus it is natural to assume that
although a given node could have a large connectivity a newly created
element will not link to that node because they have very little in
common. This feature is clearly manifested in social networks like the
WWW $-$where individuals bookmark different web pages accordingly to
their ``affinity''$-$ or the scientist citation network
\cite{rednera}. In this way, it is very unlikely to find a citation in
a condensed matter paper referring to a paper wrote by a
psychologist. Additionally, the same argument can be translated to
biological networks such as predator-prey webs or protein-protein
interaction networks.

Obviously, when $\mu$ is large enough as to dilute the first layer of
the model, we recover the BA model. The problem then consists of
determining to what extend the local preferential attachment will give
the same results, or in other words, does the knowledge of the entire
network substantially contribute to the properties observed in
the BA networks?


\begin{figure}[t]
\begin{center}
\epsfig{file=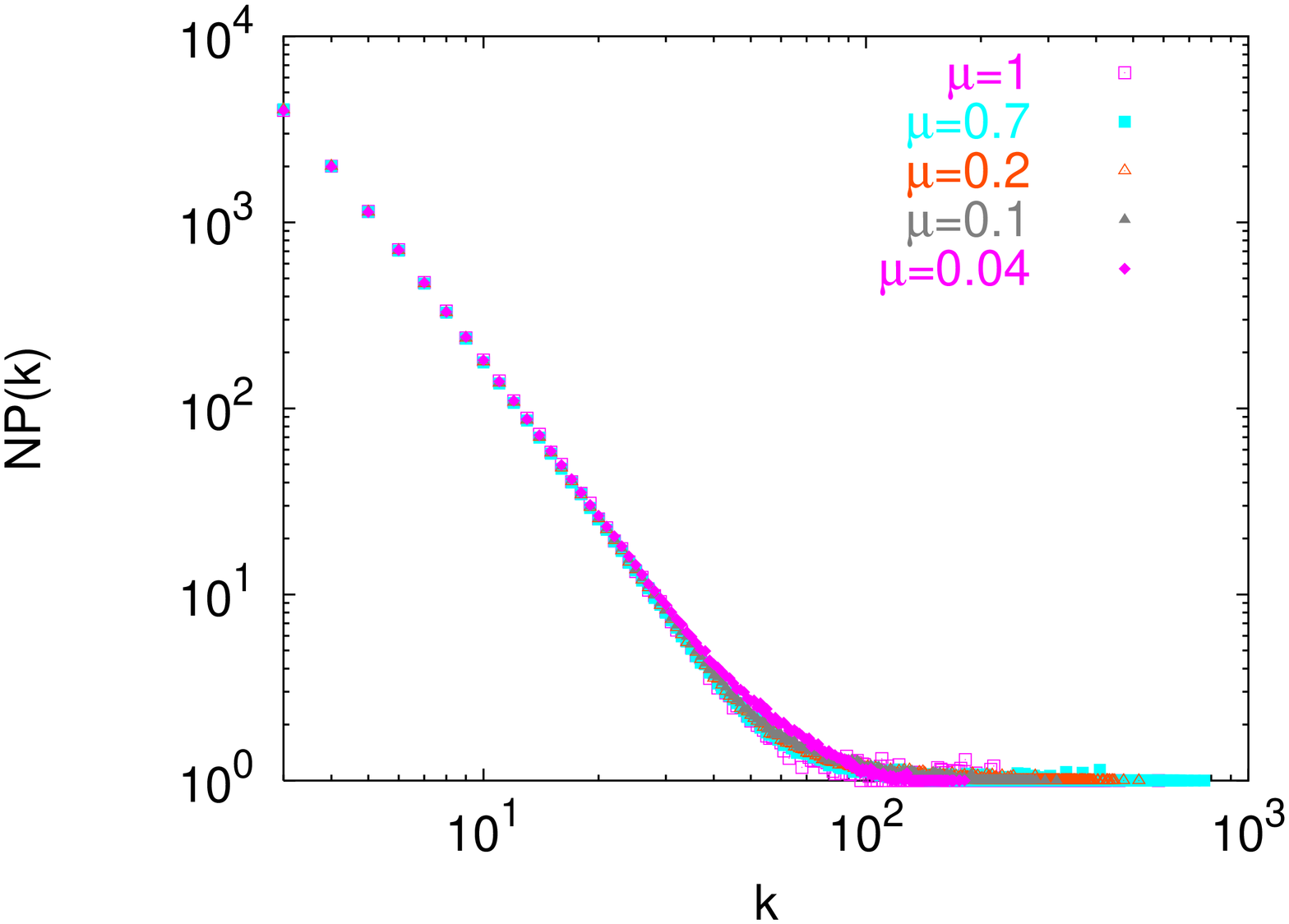,width=2.3in,angle=-90,clip=1}
\end{center}
\caption{Number of nodes with connectivity $k$ for different values of
$\mu$. The size of the network is $N=10^4$ nodes and $m_o=m=3$. The
  power-law distribution has an exponent equal to $3$. Note that the
  BA limit corresponds to $\mu=1$.}
\label{figure1}
\end{figure}

We have performed extensive numerical simulations of the model
described in the preceding section. In all cases, the numerical
results have been obtained after averaging over at least 500
iterations varying the system size from $10^3$ up to $1.2\times 10^4$
nodes. We first generate the BA network by setting the parameter
$\mu$ to its maximum value such that the preferential attachment
applies to the entire set of nodes and then tune $\mu$ in order to
systematically reduce its value and therefore the size of the set $A$
to which the second choice Eq.\ (\ref{eq1}) is applied. 

Figure\ \ref{figure1} shows the number of nodes with connectivity $k$
for several values of $\mu$. It turns out that irrespective of the
range to which the preferential attachment is applied the stationary
probability of having a node with connectivity $k$ is the same as for
the BA model, namely, $P_k\sim k^{-\gamma}$ with $\gamma\approx 3$
. This result could be intuitively understood by noting that the rules
for the network generation has been changed only at a local level, but
seeing from a global perspective the average properties should not
change radically. To realize this point, think of the network as being
made up of different small components, as given by the affinity
constraint, each of which is constructed following the BA
algorithm. It is then clear that for large system sizes, each graph
will follow the power law distribution $P_k\sim k^{-3}$ and so will
be for the entire network.

\begin{figure}[t]
\begin{center}
\epsfig{file=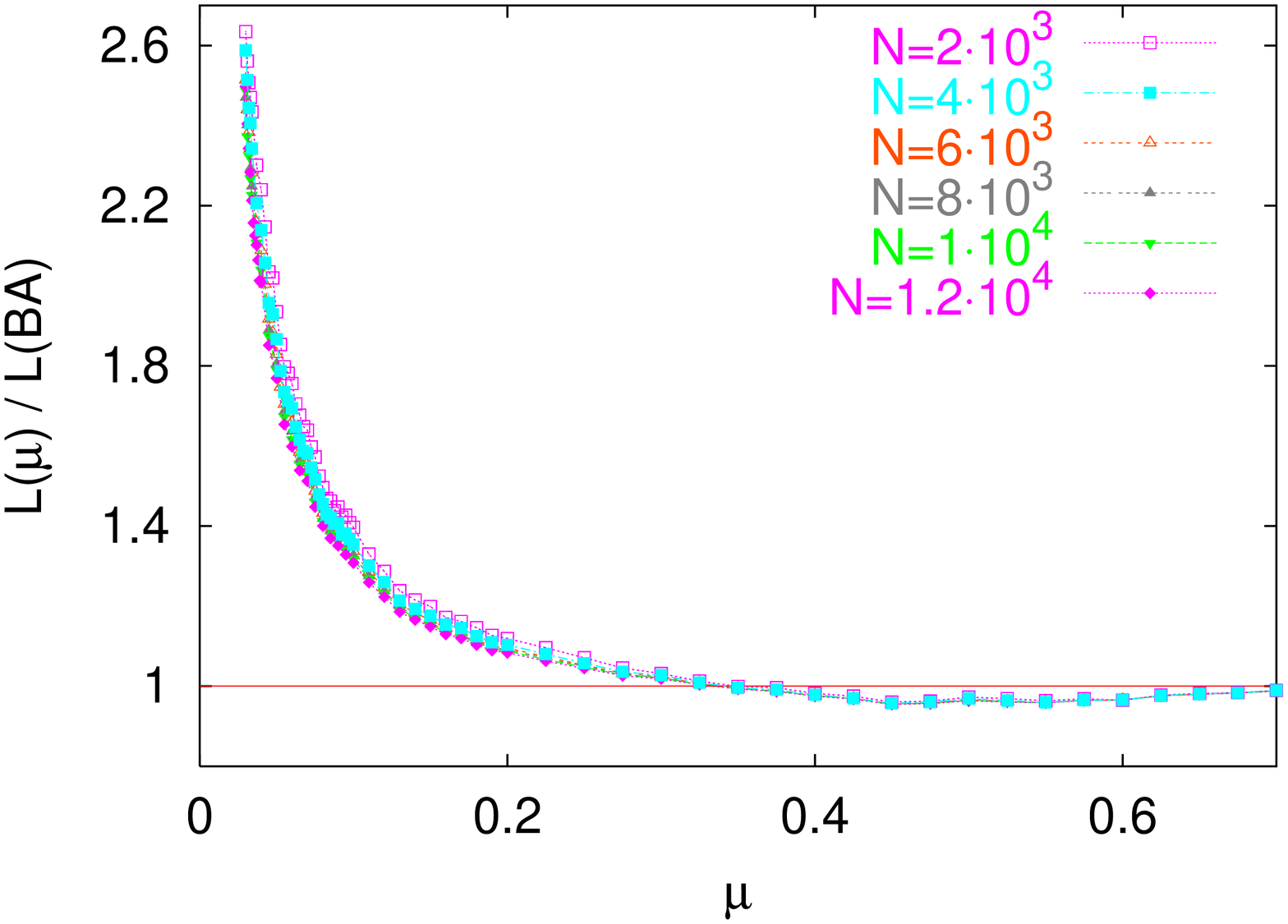,width=2.3in,angle=-90,clip=1}
\end{center}
\caption{Ratio between the average shortest path length for different
  $\mu$ values, $L(\mu)$, and that of the BA network ($L(1)$) for
  several system sizes. The horizontal line marks the BA limit. A
  transition from graphs fulfilling the small-world property to a
  regime in which networks break down in many small pieces rising the
  value of $L(\mu)$ is observed. See the text for further details.}
\label{figure2}
\end{figure}

The above argument applies only to average global properties, but
there is nothing that guarantees {\em a priori} that the components of
the network will link together in such a way that other properties
will not be affected. This is the case of the average shortest path
length $L$. The average shortest path length of a graph is defined as
the minimum number of nodes one has to pass by to go from one node of
the network to another randomly chosen node averaged over all possible
pairs of nodes. Complex networks show the noticeable property, known
as small-world property, that the average path length increases only
with the logarithm of its size. We expect that for high values of
$\mu$ the network is composed by a unique giant component and no
fragmentation arises. When the range to which the affinity criterion
is applied decreases, the network will gradually loose its compactness
and will stretch approaching a one-dimensional structure with some
small components. Further reduction of $\mu$ provokes the break down
of the network in many isolated clusters.

\begin{figure}[t]
\begin{center}
\epsfig{file=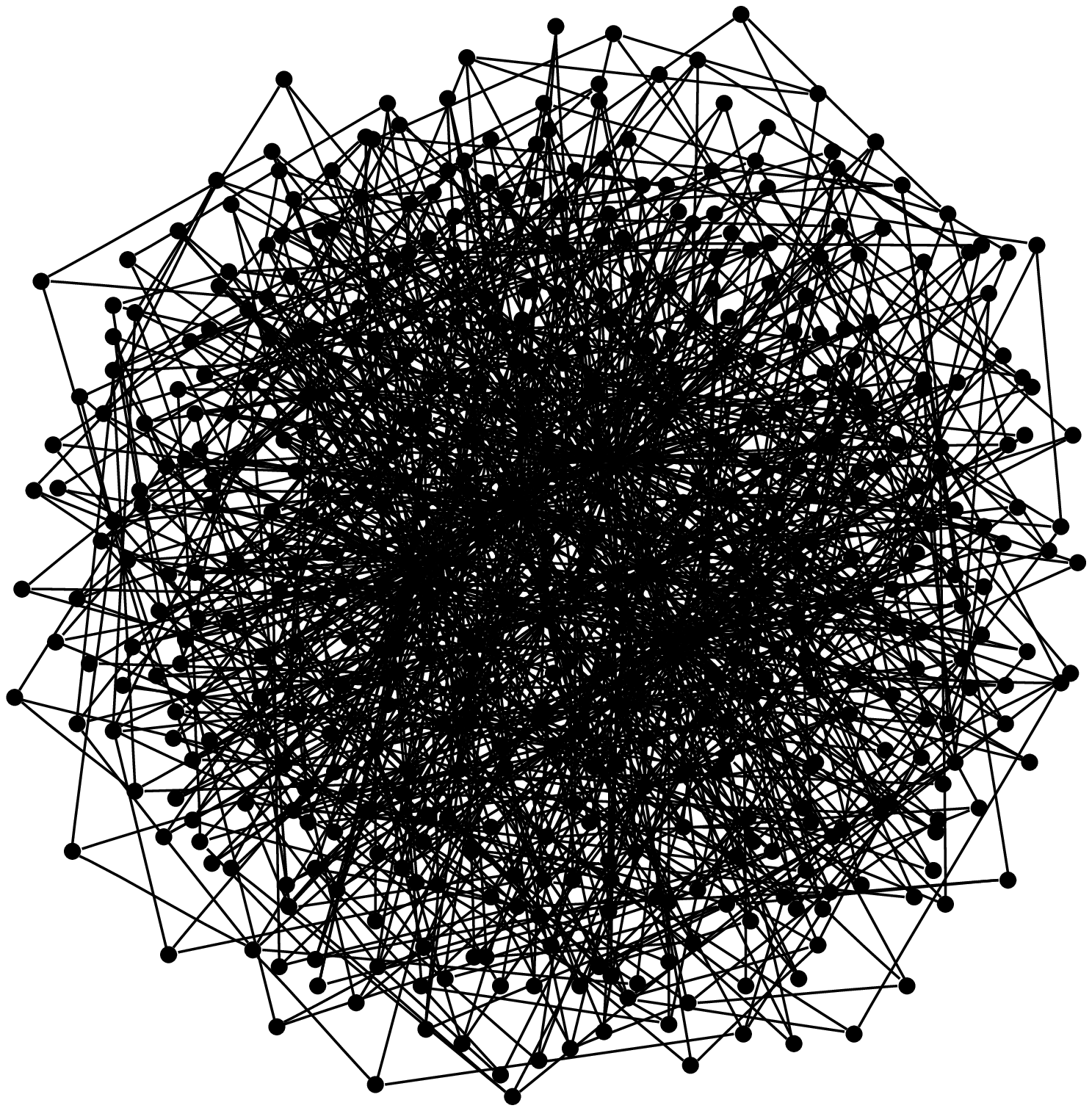,width=1.5in,angle=-90,clip=1}
\epsfig{file=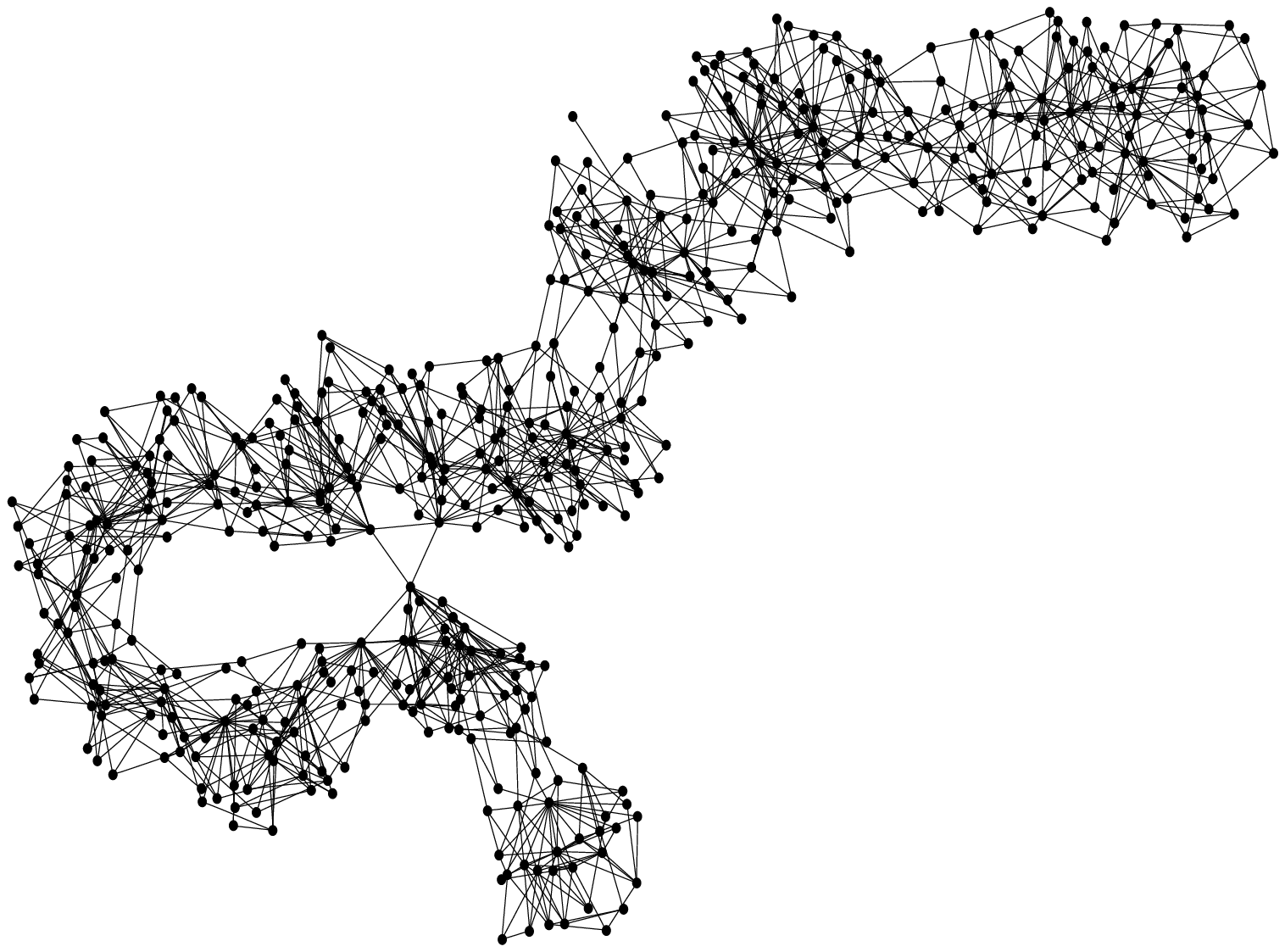,width=2.0in,angle=-90,clip=1}
\end{center}
\caption{Graph representations of two networks produced with different
  values of $\mu$. From left to right, $\mu=1$, and $\mu=0.04$. Each
  network is made up of $N=500$ nodes.}
\label{figure3}
\end{figure}

Figures\ \ref{figure2} and\ \ref{figure3} substantiate this
picture. Figure\ \ref{figure2} represents the ratio between the
average path length obtained for different values of $\mu$ and that of
the BA network, for several system sizes. As $\mu$ restricts the PA
range, the network undergoes a transition characterized by a growth of
$L(\mu)$ an eventually becomes fragmented giving rise to an infinite
shortest path length. We note here that although the results shown in
the figure have been obtained for a uniform distribution of affinity
values $a_i$, the qualitative behavior does not change for other
probability distributions and only the value at which the transition
is observed slightly shifts to the right. The shape of the network as
the parameter $\mu$ is varied can be observed in Fig.\ \ref{figure3},
where we have represented how the network looks like for the limiting
values of $\mu$. It is clear that when the PA range reduces too much
the structure of the network radically changes while keeping the same
degree distribution.

\begin{figure}[t]
\begin{center}
\epsfig{file=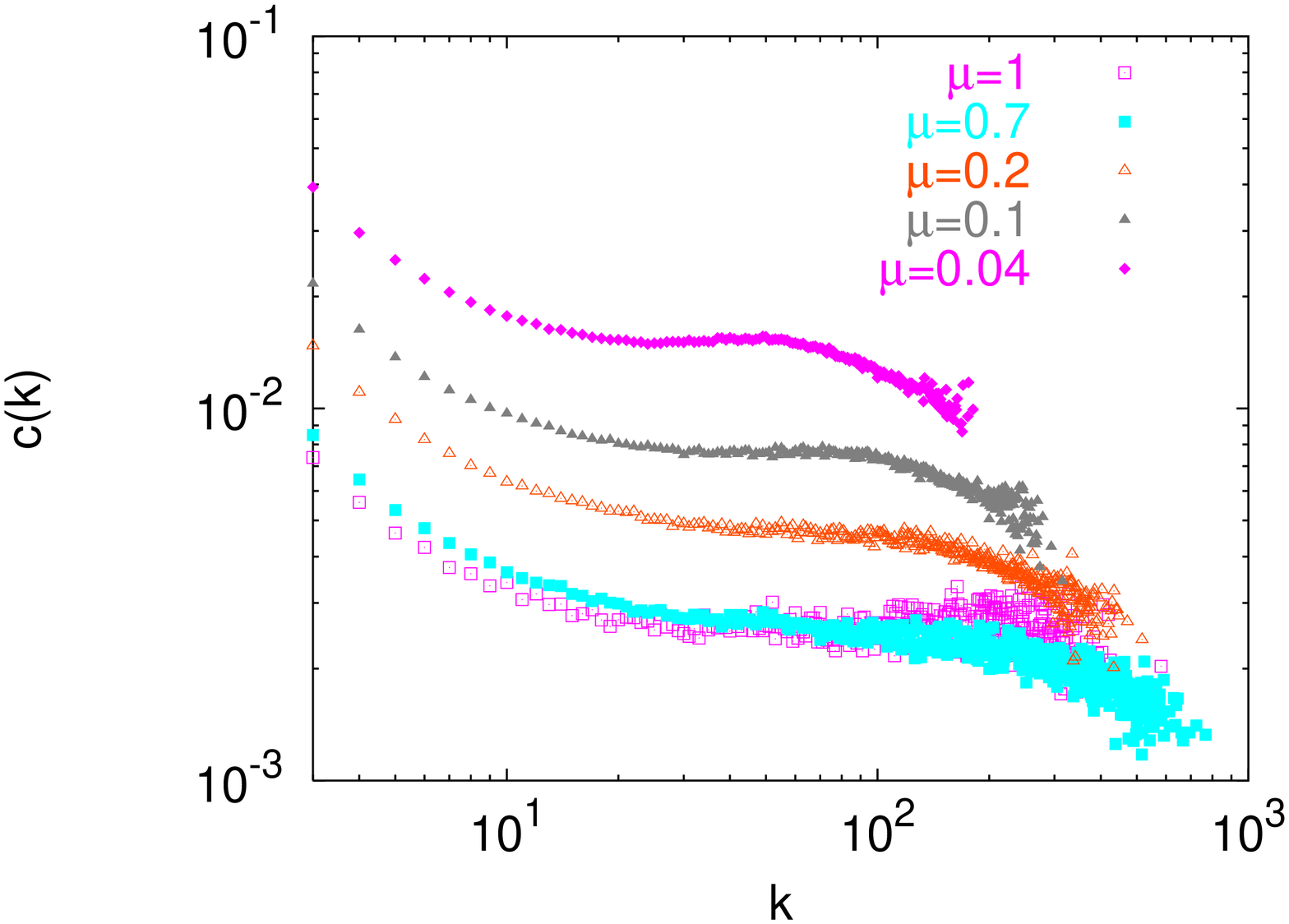,width=2.3in,angle=-90,clip=1}
\end{center}
\caption{Average clustering coefficient $c_k$ of nodes with degree $k$
for five different values of the parameter $\mu$. Note that as $\mu$
decreases, the clustering coefficient departs from the BA limit
($\mu=1$). The parameters used for the generation of the networks are
as of fig.\ \ref{figure1}.}
\label{figure4}
\end{figure}

We now focus our attention on other properties with a local
character. This is the case of the clustering coefficient $c_i$. The
clustering coefficient of a node $i$ is defined as the ratio between
the number of edges $e_i$ among the $k_i$ neighbors of $i$ and its
maximum possible value, $k_i(k_i-1)/2$, {\em i.e.},
$c_i=\frac{2e_i}{k_i(k_i-1)}$. In this way, the average clustering
coefficient, $c$ is given by the average of $c_i$ over all nodes of
the network. The clustering coefficient is of local character as it
gives the probability that two nodes with a common neighbor are also
linked together. Thus, it is expected that this magnitude, in our
model, depends on the affinity of each node and the range of
preferential attachment given by $\mu$. Figure\ \ref{figure4} shows
the average clustering coefficient of nodes with a given connectivity
$k$, for different values of the parameter $\mu$. The BA limit
exhibits almost no correlations with the degree $k$ of the vertices
and the smallest value for the clustering coefficient. As $\mu$ is
reduced, the first selection of nodes by their affinity values plays a
more dominant role contributing to the rising of $c_i$ for small and
large connectivities. Near the transition, $\mu\sim0.04$, the average
coefficient is about one order of magnitude greater than that of the BA
network. 

\begin{figure}[t]
\begin{center}
\epsfig{file=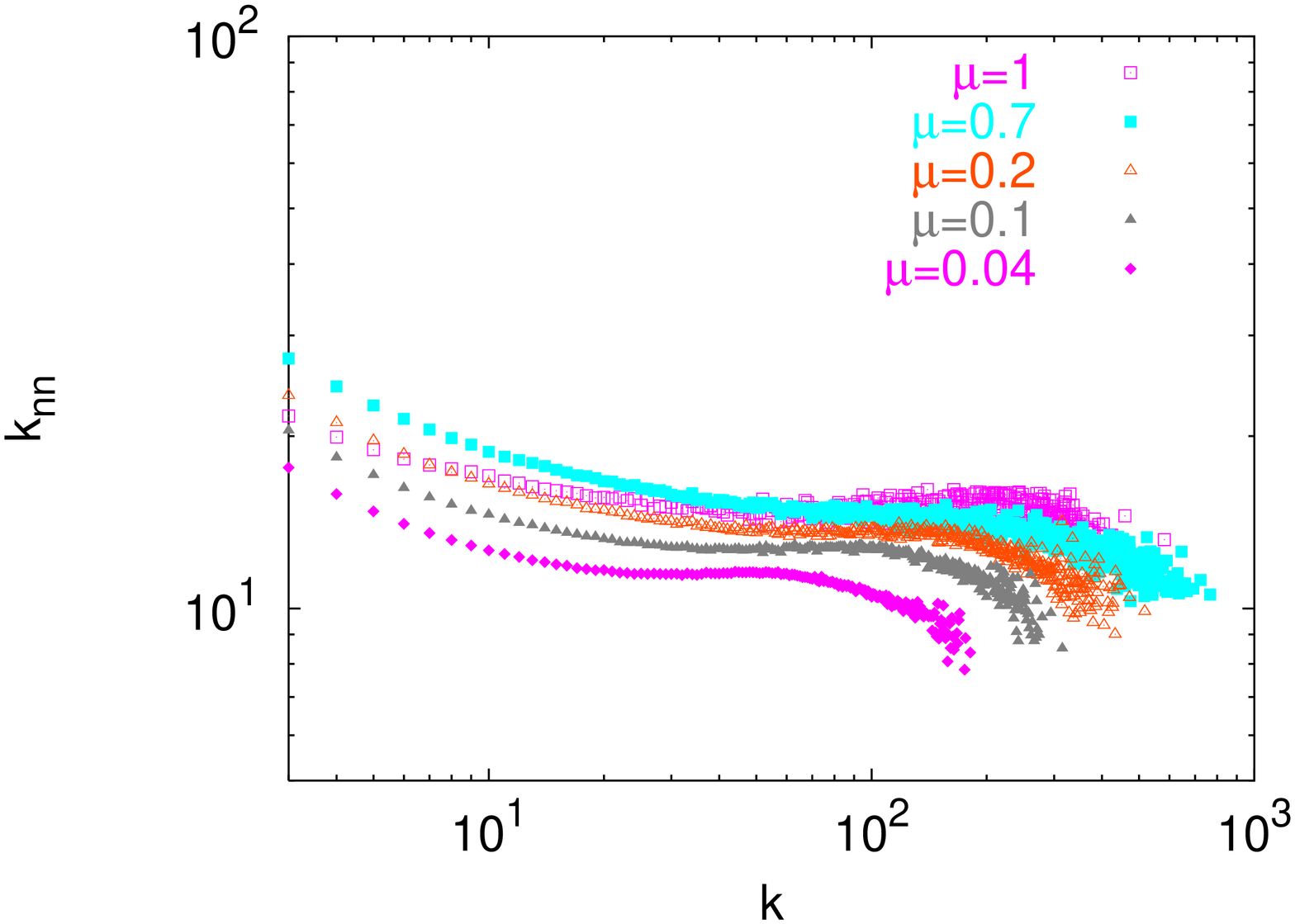,width=2.3in,angle=-90,clip=1}
\end{center}
\caption{Average nearest neighbor connectivity $k_{nn}$ against $k$
  for several values of $\mu$. Results are averaged over 100 network
  realizations for each $\mu$ value. Other parameters are as of fig.\
  \ref{figure1}.}
\label{figure5}
\end{figure}

Recently, a lot of attention has been given to network motifs
\cite{milo2,cap}, which can be defined as graph components that are
observed in a given network more frequently than in a completely
random graph with identical $P_k$. Triangles and rectangular loops are
among these graph components, also known as cycles. They are important
because they express the degree of redundancy and multiplicity of
paths among nodes in the topology of the network. The results obtained
for $c_k$ indicate that as the region where the PA applies is
reduced, the number of cycles increases and non-random correlations
arise. This is illustrated in Fig.\ \ref{figure5}, where the average
nearest neighbor degree, $k_{nn}(k)$ of a node with connectivity $k$
is depicted. While the BA model exhibits no correlations, it is
manifested the tendency that networks generated with small values of
$\mu$ display disassortative mixing at both ends of the connectivity
range.


In this paper, we have studied a version of the Barab\'asi and Albert
scale-free model that allows to tune the range to which the
preferential attachment is applied. The model considers that all nodes
are different such that they are in principle unable to link to very
distinct nodes. By introducing an affinity selection before applying
the preferential attachment rule, we tested whether or not the
knowledge of the entire network is an essential requisite to get
scale-free networks. Our results seem to support the idea that having
at least some degree of preferential attachment is enough to get an SF
growing network. We found that the connectivity distribution is not
affected by the affinity constraints while the network is unable to
link together if the tolerance range is reduced too much. On the other
hand, local properties such as the clustering coefficient do change
and reach values higher than those expected for random networks with
the same degree distribution. However, the growth of the clustering
coefficient due to the differentiation of nodes produces at the same
time a rising in the value of the average shortest path
length. Eventually the network breaks down in small pieces and looses
its small-world character.

Finally, we point out that although the values found for several
magnitudes can not be directly associated with real data, there are
some regions of the parameter space $\mu$ where non-trivial properties
arise. In this sense, it would be interesting to perform the same
analysis in more realistic growing network models looking for more
similarities with real-world networks. For example, the exponent of
the connectivity distribution can be tuned to small values by
incorporating the first level of selection of the present model in the
generalized BA model \cite{doro}, which is known to give arbitrary
$\gamma$ values in the interval $(2,3)$.

\begin{acknowledgments}
The authors thank F.\ Falo, J.\ L.\ Garc\'{\i}a-Palacios, L.\ M.\
Flor\'{\i}a and A.\ F.\ Pacheco for helpful comments and
discussions. J.\ G-G\ acknowledges financial support of the CSIC
through an I3P-BPD2002-1 grant. Y.\ M.\ is supported by the
Secretar\'{\i}a de Estado de Educaci\'on y Universidades (Spain,
SB2000-0357). This work has been partially supported by the Spanish
DGICYT project BFM2002-01798.
\end{acknowledgments}


\begin{thebibliography}{99}

\bibitem{pnas} M. E. J. Newman, Proc. Natl. Acad. Sci. U.S.A. {\bf
  98}, 404 (2001).

\bibitem{ref5} H. Jeong, S. P. Mason, A.-L. Barab\'asi, and
  Z. N. Oltvai, Nature (London) {\bf 411}, 41 (2001).

\bibitem{ref6} R. V. Sol\'e, and J. M. Montoya, Proc. R. Soc. London B
  {\bf 268}, 2039 (2001).

\bibitem{ref7} M. Faloutsos, P. Faloutsos, and C. Faloutsos,
  {\em Proceedings of the ACM, SIGCOMM} [Comput. Commun. Rev. {\bf
  29}, 251 (1999)]

\bibitem{ref8} G. Caldarelli, R. Marchetti, and L. Pietronero,
  Europhys. Lett. {\bf 52}, 386 (2000).

\bibitem{doro} S. N. Dorogovtsev and J. F. F. Mendes, Adv. Phys. {\bf
51}, 1079 (2002).

\bibitem{bara02} R. Albert and A.-L. Barab\'{a}si,
Rev. Mod. Phys. {\bf 74}, 47 (2002).

\bibitem{book1} S. N. Dorogovtsev and J. F. F. Mendes, {\it
Evolution of Networks. From Biological Nets to the Internet and the
WWW}, Oxford University Press, Oxford, U.K., (2003).

\bibitem{book2} {\it Handbook of Graphs and Networks}, Edited by
S. Bornholdt and H. G. Schuster, Wiley-VCH, Germany, 2003.

\bibitem{av03} A. V\'{a}zquez, and Y. Moreno, Phys. Rev. E {\bf 67},
	015101(R) (2003).

\bibitem{moreno03} Y. Moreno, J. B. G\'omez, and A. F. Pacheco,
Phys. Rev. E {\bf 68}, 035103(R) (2003).

\bibitem{bar99} A.-L. Barab\'{a}si, and R. Albert, Science {\bf 286},
509 (1999).

\bibitem{bar99a}A.-L. Barab\'{a}si, R. Albert, and H. Jeong, Physica A
{\bf 272}, 173 (1999).

\bibitem{amaral} S. Mossa, M. Barthelemy, H. E. Stanley, and
L. A. N. Amaral, Phys. Rev. Lett. {\bf 88}, 138701 (2002).

\bibitem{redner} P. L. Krapivsky, and S. Redner, Phys. Rev. E {\bf
63}, 066123 (2001).

\bibitem{cald} G. Caldarelli, A. Capocci, P. De Los Rios, and
  M. A. Mun\~{n}oz, Phys. Rev. Lett. {\bf 89}, 258702 (2002).

\bibitem{bia1} G. Bianconi, A. -L. Barab\'asi, Europhys. Lett. {\bf
  54}, 436 (2001).

\bibitem{note2} In case that the number of elements in the set $A$ is
smaller than $m$ we just add a link to all nodes in $A$ without
applying the PA rule.

\bibitem{rednera} S. Redner, Eur. Phys. J. B {\bf 4}, 131 (1998). 

\bibitem{milo2} R. Milo, S. Shen-Orr, S. Itzkovitz, N. Kashtan,
D. Chklovskii and U. Alon, Science {\bf 298}, 824 (2002).

\bibitem{cap} G. Bianconi and A. Capocci, Phys. Rev. Lett. {\bf 90},
078701 (2003).

\end{thebibliography}
\end{document}